\documentclass[twocolumn,showpacs,preprintnumbers,amsmath,amssymb]{revtex4}

\usepackage{graphicx}
\usepackage{dcolumn}
\usepackage{bm}

\begin{document}

\preprint{Bof}

\title{Transport properties of chemically synthesized polypyrrole thin films}

\author{C.C. Bof Bufon}\author{T. Heinzel}
\email{thomas.heinzel@uni-duesseldorf.de}
\affiliation{Heinrich-Heine-Universit\"at, Universit\"atsstr.1,
40225 D\"usseldorf, Germany}

\date{\today}

\begin{abstract}

The electronic transport in polypyrrole thin films synthesized chemically from the vapor phase is studied as a function
of temperature as well as of electric and magnetic fields. We find distinct differences in comparison to the behavior
of both polypyrrole films prepared by electrochemical growth as well as of the bulk films obtained from
conventional chemical synthesis. For small electric fields
$F$, a transition from Efros-Shklovskii variable range hopping to Arrhenius activated transport is observed at
$\mathrm{30\,\mathrm{K}}$. High electric fields induce short range hopping. The characteristic hopping distance is found to
be proportional to $1/\sqrt{F}$. The magnetoresistance $R(B)$ is independent of $F$ below a critical
magnetic field, above which $F$ counteracts the magnetic field induced localization.

\end{abstract}


\maketitle

\section{\label{sec:1}INTRODUCTION}

Polypyrrole (PPy) is a semiconducting polymer with some unusual properties compared to other organic semiconductors \cite{Skotheim1998}. The material can be synthesized both chemically and electrochemically; the electrosynthesis \cite{Yoon1994} can be performed at low polymerization potentials, allowing the use of water as solvent \cite{Asavapiriyanont1984,Rodriguez1997,Simonet1991}. Typically, such films have a rather large thickness above 1 micron, and they are highly disordered as well as crosslinked; maximum conductivities of 1600 S/cm have been reported for stretched films \cite{Yamamura1988}. In addition, PPy can be doped electrochemically to a high level, which stabilizes the material properties under ambient conditions \cite{Kanazawa1980,Diaz1981,BofBufon2005}. A wide range of applications has been demonstrated, such as gas sensors \cite{Marcos1997,Bidan1992,Malinauskas2001,Geng2006}, biosensors \cite{Cosnier2000}, or transistors \cite{Chung2005,Lee2005,Bof2006}.

The transport properties of electrochemically synthesized PPy films have been studied extensively \cite{Yoon1994,Skotheim1998}. In the insulating state that is usually present, the samples typically follow the Efros-Shklovskii (ES) variable range hopping model \cite{Shklovskii1988,Efros1975}. Such a system is assumed to be formed by metallic sites separated by tunnel barriers \cite{Menon1998,Kohlman1998}, which are
generated by spatial variations in the charge density. The electric field dependence of the conductivity was investigated by Ribo et al. \cite{Ribo1998}. More recently, Gomis and co-workers \cite{Gomis2003} reported
an electric field - induced transition from an insulating to a \emph{metallic} state which is reached when the electric field is strong enough to delocalize a small fraction ($<\mathrm{0.01\%}$) of the charges. \\

Chemical synthesis \cite{Joo2000}, on the other hand, allows film preparation on conductive as well as on insulating substrates, thereby complementing the range of applications. Chemically grown films have a pronounced egg-like morphology
\cite{Zhang2006}, markedly different than the characteristic fibers obtained by electrochemical synthesis. These films are, however, of notoriously poor quality in terms of roughness and conductivity. For example, their reported maximum room temperature conductivities are $40\,\mathrm{S/cm}$ \cite{Chakrabarti2002,Winther2004}. This may be the reason why only a very limited number of works have addressed the transport properties of this kind of PPy films. The existing studies have demonstrated that in films with thicknesses larger than $60\,\mathrm{\mu m}$, the temperature dependence $\sigma(T)$ of the conductance is consistent with three-dimensional Mott variable range hopping, i.e. $ln(\sigma)\propto T^{-1/4}$ \cite{Joo2000,Chakrabarti2002}. Up to now, chemical synthesis of PPy at reduced temperatures has not been reported, which is somewhat surprising considering the profound influence the growth temperature has on the conductivity of electrochemically synthesized films \cite{Yoon1994}.\\

Here, we present an investigation of the transport properties of PPy thin films and low roughness, prepared by a recently established scheme that allows the formation of thin films by chemical polymerization from the vapor phase \cite{Bof2006}. The differential conductance is studied as a function of electric and magnetic fields as
well as of temperature. With small electric fields applied, we find that for low temperatures, the electronic transport can be
described within the ES-variable range hopping model. Above a temperature of $30\,\mathrm{K}$, Arrhenius-type activated transport is found. As the electric field is increased, a transition to short-range hopping is observed around a characteristic electric field of $F_c=2500\,\mathrm{V/cm}$. This electric field - induced reduction of the localization is also visible in magnetotransport measurements.

\section{\label{sec:2}SAMPLE PREPARATION AND CHARACTERIZATION}

PPy films were synthesized via chemical polymerization from pyrrole vapor in solutions of $\mathrm{HCl:H_2O_2}$.
The preparation scheme follows that one described in detail in a previous publication \cite{Bof2006}, except that our
solvent was cooled, with a minimum growth temperature of $T_g=277.5\,\mathrm{K}$. At lower temperatures, the polymerization kinetics
became unacceptably slow. A thin PPy film was formed on a doped and oxidized Si substrate containing four Pt
electrodes. The film thickness is proportional to the exposure time
to the vapor and can be tuned between a few nanometers and $\approx 0.5\,\mathrm{\mu m}$. In the
present study, we focus on films of thickness $70\pm 10\,\mathrm{nm}$. We observe that besides growth at $T_g>282\,\mathrm{K}$, also water created by decomposition of $\mathrm{H_2O_2}$, an aging process
in the solution, increases the contact resistance. Therefore, fresh solution has been used for each sample.

Films have been grown at temperatures between $277.5\,\mathrm{K}$ and $300\,\mathrm{K}$ in solutions with volume
fractions varying between $\mathrm{HCl:H_2O_2 = 1:1000}$ and $5:1000$. Cyclic voltammetry measurements \cite{BofBufon2005} allow us to estimate
the doping density of all samples to $p\approx 10^{21}\,\mathrm{cm^{-3}}$. This and the fact that the doping
density depends neither on the growth temperature nor on the volume fraction of the solution indicates that the films
are in the fully charged state, corresponding to 0.25 and 0.33 holes per pyrrole ring \cite{Bredas1983,Bredas1984}. The
electronic transport has been studied in eleven
samples, all showing qualitatively identical behavior. Typical room temperature conductivities of
$10\,\mathrm{S/cm}<\sigma (300\,\mathrm{K})<20\,\mathrm{S/cm}$ are found, with no apparent systematic
dependence on $T_g$ or on the volume fraction of the solution. At temperatures below $\approx 30\,\mathrm{K}$,
however, samples grown at $T_g>282\,\mathrm{K}$ show substantially lower conductivities than samples grown below this temperature.
In the following, we focus on three samples, the parameters of which are summarized in Table \ref{tab:01}.

\begin{table}
  \centering
\begin{tabular}{|c|c|c|c|c|c|c|}
  \hline
  Sample& $T_g$ &volume& $\mathrm{\sigma(300\,\mathrm{K})}$&$\mathrm{\xi}$&$\mathrm{r(2.1\,\mathrm{K})}$ &$\mathrm{r(30\,\mathrm{K})}$\\  &[K]&fraction&[S/cm]&[nm]&[nm]&[nm]\\
  \hline
  A & 282.0 & 1:1000 & 10& 1.5 & 30 & 8\\
  B & 277.5 & 1:1000 & 12& 3.4 & 44 &12\\
  C & 277.5 & 4:1000 & 20& 8.8 & 72 &19\\
  \hline
\end{tabular}
\caption{Summary of the sample growth conditions and transport parameters. $T_g$ denotes the growth temperature, $\xi$ the
localization length and $r$ the ES-variable range hopping distance.}\label{tab:01}
\end{table}

\begin{figure}[h t b]
\centering
\includegraphics[scale=0.4]{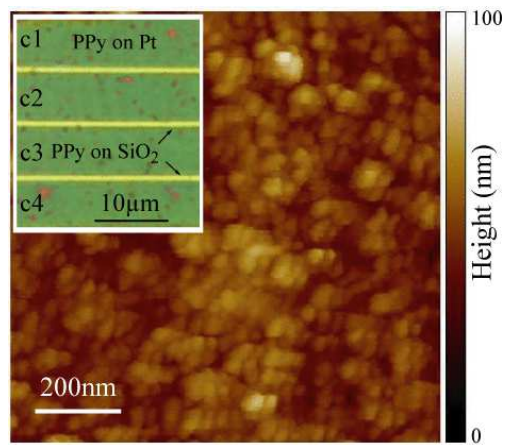}
\caption{Morphology of sample A, as measured with an atomic force microscope. The inset shows a top view of the PPy film
on top of the four Pt electrodes.} \label{fig:1}
\end{figure}

Figure \ref{fig:1} shows an atomic force microscope image of the surface of sample A and the corresponding
four-electrode device as seen in an optical microscope (inset), where the gap between the electrodes
is $\mathrm{2\,\mu m}$. The PPy film covers the 4 Pt electrodes (c1 to c4). The roughness (single standard deviation)
decreases with decreasing growth temperature, varying from 25\% of the film thickness for the samples grown at $300\,\mathrm{K}$ to
6\% for the samples grown at $277.5\,\mathrm{K}$. The film morphology shows egg-like structures well known for chemically
synthesized PPy \cite{Zhang2006}, with an average diameter of $\approx \mathrm{50\,nm}$.\\

Transport measurements have been carried out in a $\mathrm{^{4}He}$ gas flow cryostat with a temperature range between
$\mathrm{2.1\,\mathrm{K}}$ and $\mathrm{300\,\mathrm{K}}$. The cryostat is equipped with a superconducting magnet
(maximum magnetic field $\mathrm{8\,T}$). Our setup has a DC current resolution of $\approx 10\,\mathrm{pA}$. By comparing
two-probe with four probe measurements, we found negligible contact resistances for all experimental conditions
investigated. Samples that were measured at temperatures above $\mathrm{30\,\mathrm{K}}$ with applied electric fields
$F\,>\,1500\,\mathrm{V/cm}$ got permanently damaged, while below $30\,\mathrm{K}$, up to $9500\,\mathrm{V/cm}$ could be applied.

\begin{figure}[h t b]
\centering
\includegraphics[scale=0.5]{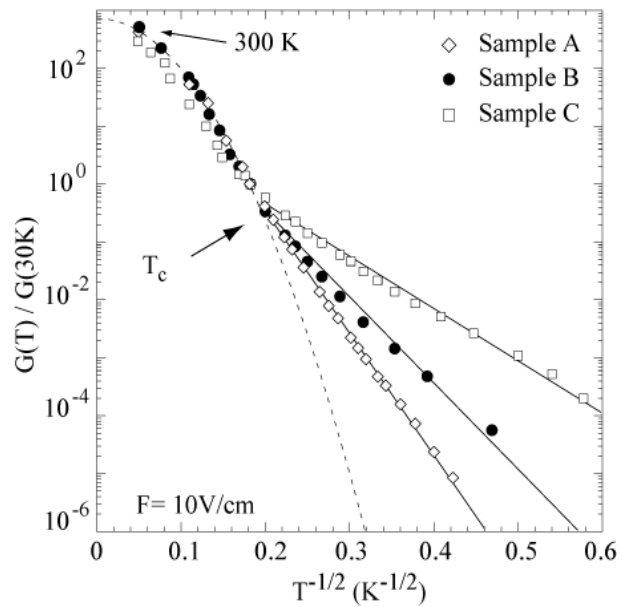}
\caption{Conductance as a function of $T^{-1/2}$ for samples A, B and C, normalized to the values
measured at $T_{c}=\mathrm{30\,K}$. The full lines are fits to ES-variable range hopping
at temperatures below $T_c$, while the dashed line represents the fit to the Arrhenius law for $T> T_{c}$ .} \label{fig:2}
\end{figure}

In Fig. \ref{fig:2}, the temperature dependence of the conductance $G(T)$ for the three samples under small electric fields ($F=10\,\mathrm{V/cm}$) is shown. Below $30\,\mathrm{K}$, $G(T)$ behaves in accordance with the standard ES - variable range hopping relation \cite{Efros1975,Shklovskii1988}, given by

\begin{equation}
G(T)\,\propto\,\exp
\left[-\left(\frac{T_0}{T}\right)^{1/2}\right]\label{eq:1}
\end{equation}
with
\begin{equation}
T_0\,=\,\frac{2.8e^2}{4\pi\varepsilon_{0}\varepsilon_{r}
k_{B}\xi}\label{eq:2}
\end{equation}

Here, $\xi$ denotes the localization length, and $\varepsilon_{r}=13.6$ is the dielectric constant of PPy \cite{Van2000,Koezuka1983}.
Below $4.6\,\mathrm{K}$, the current vanishes in our noise floor. We have fitted $G(T)$ to eq. \ref{eq:1} for temperatures below $30\,\mathrm{K}$. For sample A, $T_0\,= 2340\,\mathrm{K}$ is obtained, corresponding to $\xi\,=1.5\,\mathrm{nm}$
(see Table \ref{tab:01} for the corresponding results obtained for samples B and C). The characteristic hopping distance $r$ within the ES-variable range hopping model is given by

\begin{equation}
r\,=\,\left(\frac{{e^{2}\xi}}{4\pi
\varepsilon_{0}\varepsilon_{r} k_{B}T}\right)^{1/2}\label{eq:3}
\end{equation}

The values of $r$ at $\mathrm{2.1\,K}$ and $\mathrm{30\,K}$ are displayed in Table \ref{tab:01}. At the critical temperature $T_c = 30\,\mathrm{K}$, $G (T)$ shows a crossover from ES-variable range hopping to Arrhenius behavior, which is characterized by
\begin{equation}
G(T)\propto \exp (-U/k_{B}T){\label{eq:4}}
\end{equation}

Such a crossover is expected at a temperature where nearest neighbor hopping begins to dominate over variable range hopping \cite{Mott1974}. It has been observed in several other systems, for example in gold nanoparticle multilayers \cite{Tran2005}, or in
$\mathrm{ZnO}$ nanocrystal assemblies \cite{Roest2003}.

For nearest neighbor hopping, the activation energy $U$ is given by \cite{Yu2004}

\begin{equation}
U\,=\,\frac{\xi}{8d}{k_BT_0}{\label{eq:5}}
\end{equation}

where $d$ denotes the nearest neighbor distance. From fits of eq.\eqref{eq:4} to the data, the activation energy is found to be
$U\,=(\,16\,\pm 1)\,\mathrm{meV}$ for all samples. Via eq. \ref{eq:5},
this corresponds to $d\,=(\,2\,\pm 0.5)\,\mathrm{nm}$.\\
\begin{figure}[h t b]
\centering
\includegraphics[scale=0.5]{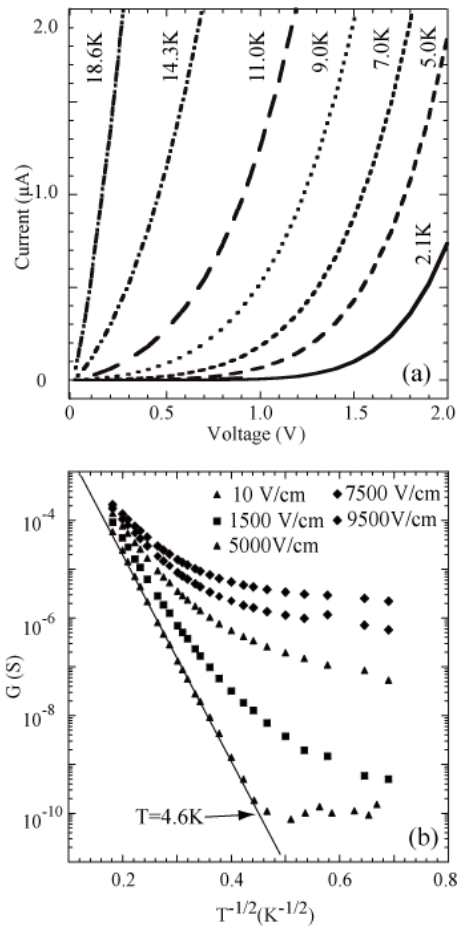}
\caption{(a) Measured IV characteristics of sample A  at different temperatures. (b) The temperature dependence of
$G(T)\,=\,dI/dV(T)$ , for various electric fields $F$. The straight line is the fit of $G(T)$ at
$F=10\,\mathrm{V/cm}$ for sufficiently large temperatures, i.e. for conductances above the noise floor,
to the Efros-Shklovskii model, see Fig. \ref{fig:2}.} \label{fig:3}
\end{figure}

We proceed by discussing the effect of high electric fields. Fig. \ref{fig:3} (a) shows the IV traces as a function of the temperature for sample A. All samples exhibit a non-linear IV characteristics below $\mathrm{90\,K}$. In Fig. \ref{fig:3} (b), the differential
conductance $\mathrm{G(T)}$ is shown for sample A for various electric fields $F$. Clearly, the electric field has a profund effect on the conductance; it can induce changes of $G$ over more than four orders of magnitude at low temperatures over the accessible electric filed range, and strong deviations from eq. \eqref{eq:1} occur: the shape of $G(T)$ becomes strongly dependent on $F$, while the temperature dependence of $G$ decreases with increasing electric field. Our data can be interpreted within the extension of the ES - variable range hopping model to significant electric fields: it has been argued that for samples in the ES-variable range hopping regime, the electric field reduces the Coulomb energy over a characteristic hopping distance
$r$ by $eFr$ \cite{Dvurechenskii1988}, which for intermediate electric fields results in a modified conductivity given by

\begin{equation}
G(F,T)\,\propto\,\exp
\left(-2\frac{r}{\xi}\,-\,\frac{\xi}{2.8r}\frac{T_0}{T}\,+\,\frac{e
Fr}{k_BT}\right)\label{eq:6}
\end{equation}

\begin{figure}[h t b]
\centering
\includegraphics[scale=0.5]{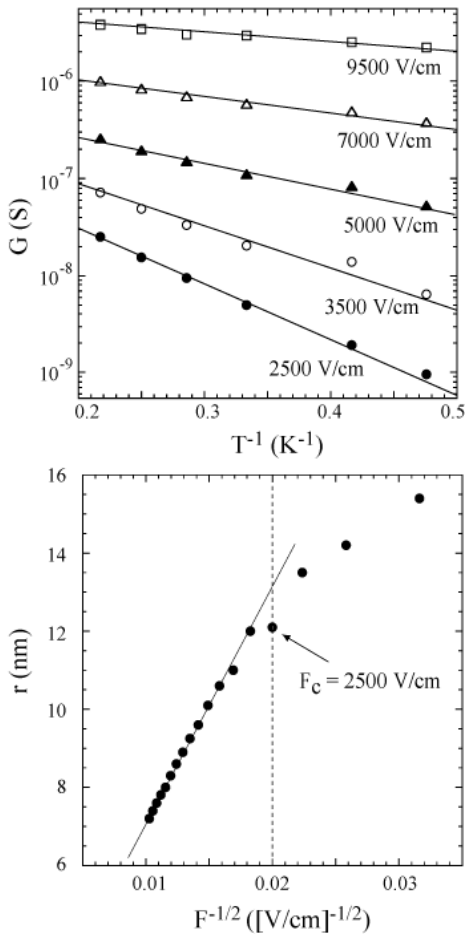}
\caption{(a) Differential conductance of sample A (symbols)
and the corresponding fits to eq. \eqref{eq:6} in the range
$eF \xi \gtrsim k_BT$. (b) The electric field
dependence of the fitted values for $r$ follow a $1\sqrt{F}$ dependence ( full line)
above the critical field $F_c=2500\,\mathrm{V/cm}$ (dashed line).} \label{fig:4}
\end{figure}

This intermediate regime is defined by $F_0=k_BT/eL\leq F\leq 2k_BT/(e\xi)=F_c$, where $F_0$ and $F_c$ denote a lower and
upper critical electric field, respectively. $L$ is a length parameter of the order of the maximum hopping length, the
exact numerical value of which is under controversial discussion \cite{Hill1971,Pollak1976,Shklovskii1976}.
In this intermediate regime, one should expect that $r$ depends on both temperature and electric field, and various models
for this relation have been discussed \cite{Hill1971,Pollak1976,Shklovskii1976}. However, we find that for
$F\gtrsim 2500\,\mathrm{V/cm}\approx F_c(5\,\mathrm{K})$ and for temperatures below $5\,\mathrm{K}$, $ln(G)$ shows a  $1/T$
dependence to high accuracy, see Fig. \ref{fig:4}(a). According to eq. \eqref{eq:6}, this would be expected for a
temperature independent hopping distance. This observation implies that in this regime, $r$ is mostly determined by the
electric field in our samples ($r=r(F)$) and depends only weakly on temperature. We have therefore fitted $G(T)$ for
different electric fields in this regime under the assumption of a temperature-independent hopping distance, using $r(F)$ as
a fit parameter.

In Fig. \ref{fig:4}(b), these values are shown as a function of $F^{-1/2}$. It is observed that for electric fields above
$\approx 2500\,\mathrm{V/cm}$, $r(F)$ becomes proportional to $F^{-1/2}$, with a slope of $dr/d(F^{-1/2})= 6.13\times 10^{-6}\,\mathrm{(Vm)^{1/2}}$. For lower fields, $r(F)$ deviates from this relation and the fits become worse, probably because in this range, $r$ develops a significant temperature dependence. This observation is in qualitative agreement with the model of Dvurechenskii et al. \cite{Dvurechenskii1988}, where it has been argued that in high electric fields, i.e. for $F> k_b T/(e\xi)$, the energy gained by the electric field overcomes the activation energy such that the transport becomes activationless, which means that the last two terms in the exponent of eq. \eqref{eq:6} cancel each other, and a temperature-independent conductivity evolves with an electric field dependence given by

\begin{equation}
G(F)\,\propto\,\exp
\left[-\frac{2r(F)}{\xi}\right]\label{eq:7}
\end{equation}

with
\begin{equation}
r(F)=\sqrt{\frac{e}{4\pi \epsilon \epsilon_0}}\frac{1}{\sqrt{F}}
\end{equation}
This model predicts a slope of $dr/d(F^{-1/2})= 1.03\times 10^{-5}\,\mathrm{(Vm)^{1/2}}$, in rough agreement with (i.e. 70\% larger than) our value determined from the experiment. We speculate that either the dielectric constant increases in high electric fields, or the electric fields accessible in our experiment are not high enough for eq. \eqref{eq:7} to hold, even though the same functional dependence is found. We note that similar discrepancies have been reported in the literature for other systems \cite{Yu2004}.

The behavior of $G(T)$ in large electric fields could be also attributed to Joule heating, which would be most prominent at low temperatures under high electric fields and cause a saturation of $G$ as the temperature decreases. From the heat generated at $F=9500\,\mathrm{V/cm}$ and the thermal conductivity of the oxide layer underneath the PPy film, we estimate the maximum increase of the sample temperature by Joule heating to $\Delta T \sim 0.5\,\mathrm{K}$. Therefore, the Joule heating effect can be neglected in our experiments.\\

\begin{figure}[h t b]
\centering
\includegraphics[scale=0.5]{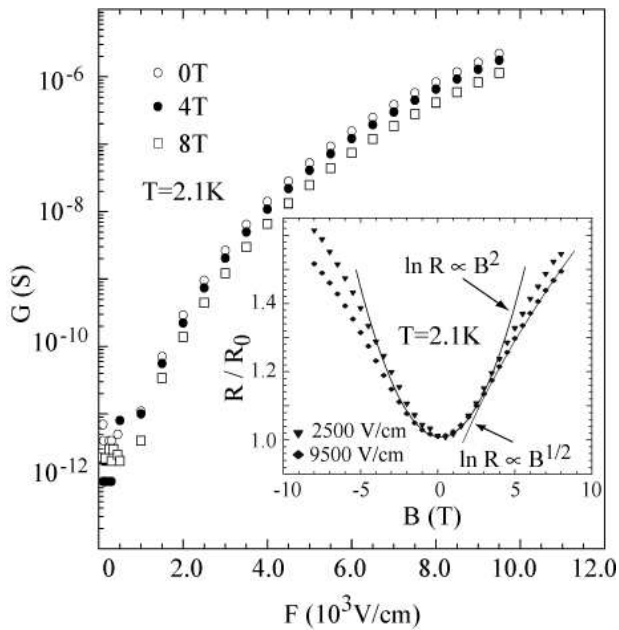}
\caption{Effect of a magnetic field on the $F$ dependence of the differential conductance as measured on sample A. Inset: the magnetoresistance as measured on this sample at $F=2500\,\mathrm{V/cm}$ and $9500\,\mathrm{V/cm}$.
The data are normalized to their value at $\mathrm{B}=0$.} \label{fig:5}
\end{figure}

Fig. \ref{fig:5} shows the differential conductance at $2.1\,\mathrm{K}$ as a function of $F$ for various magnetic fields, taking sample A as an example. For all electric fields, the magnetic field tends to localize the charges, thus counteracting the electric-field induced delocalization. In the inset of Fig. \ref{fig:5}, the magnetoconductance is shown for different electric fields. For $B\lessapprox 4\,\mathrm{T}$, $R(B)$ is independent of $F$ within experimental accuracy. At $B\approx 4 \,\mathrm{T}$, the curvature of $R(B)$ switches from positive to negative while for larger magnetic fields, the normalized resistance change induced by $B$ becomes weaker as $F$ increases.

To the best of our knowledge, the magnetoresistance of a disordered medium in the variable range hopping regime under high electric fields has not been studied up to now. The theory of magnetotransport in such systems under negligible electric fields, however, is well established \cite{Shklovskii1988,Shklovskii1972}, and in the following we use it to interpret our measurements. This seems well justified for the low magnetic field regime where only a very small influence of the electric field on $R(B)$ is observed. According to Shklovskii and Efros \cite{Shklovskii1988}, the magnetic confinement reduces the wave function overlap between sites. Within a percolation model, this causes the resistance to increase exponentially as $B$ increases according to

\begin{equation}
R(B)\,\propto exp [A(T)B^{2}]\label{eq:8}
\end{equation}

where

\begin{equation}
A(T)\,=\,0.036\left(\frac{e \xi^2}{\hbar}\right)^2\left(\frac{T_0}{T}\right)^{3/2}\label{eq:9}
\end{equation}

Below $\mathrm{4\,T}$, eq. \ref{eq:8} fits our data well, with the fit parameters $A(2.1\,\mathrm{K})=\mathrm{0.017\,T^{-2}}$ and $\mathrm{0.014\,T^{-2}}$ for the measurements at $F=\mathrm{2500\,V/cm}$ and $F=\mathrm{9500\,V/cm}$, respectively. From eq. \eqref{eq:9}, we expect $A(2.1\,\mathrm{K})\approx\, 0.015\,\mathrm{T^{-2}}$ for sample A in good agreement with the fit parameters. For $B\gtrapprox 4\,\mathrm{T}$, deviations from eq. \eqref{eq:8} are observed. Theory \cite{Shklovskii1988,Shklovskii1972} predicts that for negligible electric fields and $B\gg B_c=n^{1/3}\hbar/e\xi$ ($n$ denotes the density of impurities), the magnetoresistance has the form

\begin{equation}
R(B)\,\propto exp[\sqrt{B/B_c}]\label{eq:10}
\end{equation}

which has been observed in several experiments \cite{Yoon1994,Kahlert1976}. Our data are fitted well by eq. \eqref{eq:10}
for $B>4\,\mathrm{T}$. The measurements indicate a value of $B_c\approx 4\,\mathrm{T}$. This would correspond to an impurity density of $n=7 \times 10^{14}\,\mathrm{cm^{-3}}$, i.e. an average distance between impurity sites of $110\,\mathrm{nm}$, much larger than the characteristic hopping distance, a result which cannot be interpreted in a straightforward way. We note that even though our data can be described qualitatively by eq. \eqref{eq:10}, its application is questionable considering the significant effect $F$ has on $R(B)$ in this regime and the fact that $B$ is not large compared to $B_c$. Also, a behavior according to eq. \eqref{eq:10} for $B\approx B_c$, i.e., outside the range of validity of eq. \eqref{eq:10}, has been observed  in other experiments, see, e.g. \cite{Kahlert1976}. Additional theoretical studies are required to understand this regime in more detail.

\section{\label{sec:6}SUMMARY AND CONCLUSIONS}

The electronic transport properties of thin, chemically grown polypyrrole films have been investigated. In contrast to previous experiments performed on much thicker films, we find that Efros-Shklovskii variable range hopping dominates at temperatures below $30\,\mathrm{K}$, and Arrhenius activated transport is observed at higher temperatures. Nonlinearities in the current-voltage characteristics are found, which can be interpreted within an extension of the Efros-Shklovskii model to high electric fields. In agreement with this model, we observe that the hopping distance is proportional to $1/\sqrt{F}$ above a characteristic field of $\approx 2500\,\mathrm{V/cm}$ in our system, which marks a transition to short range hopping. Moreover,  for small magnetic fields, the measured magnetoresistance is in quantitative agreement with the Shklovskii-Efros description of magnetic field induced localization in disordered media. At large magnetic fields, we observe an electric field induced reduction of the localization which requires further theoretical study. \\

Financial support by the \emph{Heinrich-Heine-Universit\"at
D\"usseldorf} is gratefully acknowledged. The authors acknowledge fruitful discussions with D. Basko.

\newpage

\end{document}